\documentstyle[tighten,eqsecnum,aps,prd,floats]{revtex}
\input{psfig}
\begin{document}
\draft

\preprint{\vbox{\baselineskip=12pt
\rightline{}
\rightline{gr-qc/0011099}
}}

\title{Radiation Pressure Induced Instabilities in Laser
Interferometric Detectors of Gravitational Waves}

\author{A. Pai $^{1}$ \and S. V. Dhurandhar $^{1,2}$ \and P. Hello $^{3}$ \and J-Y. Vinet $^{3}$}

\address{$^{1}$ Inter-University Centre for Astronomy and Astrophysics, Post Bag 4, Ganeshkhind, Pune 411007, India.}
\address{$^{2}$ Dept. of Physics and Astronomy, UWCC, PO Box 913, Cardiff CF2 3YB, Cardiff, U.K.}
\address{$^{3}$ Groupe Virgo, Laboratoire de l'Acc\'el\'erateur Lin\'eaire, Bat. 200, Centre d'Orsay, 91405 Orsay, France.}

\maketitle
 
\begin{abstract}

The large scale interferometric gravitational wave detectors consist of Fabry-Perot cavities operating at very high powers ranging from tens of kW to MW for next generations. The high powers may result in several nonlinear effects which
would affect the performance of the detector. In this paper, we investigate the effects of radiation pressure, 
which tend to displace the mirrors from their resonant
position resulting in the detuning of the cavity. We
observe a remarkable effect, namely, that the freely hanging mirrors gain
energy continuously and swing with increasing 
amplitude. It is found that the `time delay',
that is, the time taken for the field to adjust to its instantaneous
equilibrium value, when the mirrors are in 
motion, is responsible for this effect. This effect is likely to be important 
in the optimal operation of the full-scale interferometers such as VIRGO and LIGO.

\end{abstract}


\section{Introduction}
The general theory of relativity predicts the existence of
gravitational waves. Since gravity couples very weakly to matter,
highly
sensitive detectors are required to detect gravitational
waves. 
Over the next decade several large-scale interferometric gravitational wave
detectors will come on-line. These include the LIGO, composed of two 
interferometric detectors situated in the United
States each with baselines of 4 km, VIRGO, an Italian/French project
located near Pisa with a baseline of 3 km, GEO600, a British/German
interferometer under construction near Hannover with a baseline of 600 m,
TAMA in Japan, a medium-scale laser interferometer with a baseline of 300 m and with funding 
approval AIGO500, the proposed 500 m project
sponsored by ACIGA \cite{abr,bra,geo,tsu,san}. 
The large scale interferometers will use Fabry-Perot cavities and the
ground based detectors will have arm lengths of few kilometers.
  There are several noise sources which plague the detector. 
Amongst them, the photon shot noise is dominant at high frequencies. It
is reduced by increasing the amount of power of the
laser source, as the noise is inversely proportional to the square root
of the power. Therefore the cavities envisaged will operate with 
very high powers in their arms, tens of kiloWatts for initial detectors and perhaps powers as high as
megaWatts in planned advanced detectors. The high power stored 
in the cavities can generate a number of nonlinear effects which would adversely affect the operation of the optical cavity.  
Here, we look into one such effect, namely, the dynamics of mirrors 
under the radiation
pressure force. In earlier literature, we and others had investigated the
thermo-elastic deformation of the mirrors due to the absorption of the
power in the coatings and performed a longitudinal analysis of the
cavity \cite{svd,win,hel1,hel2,hel3}. Following these investigations, we studied the 
effects of radiation pressure in the
cavity, in the regime when the displacement of the mirror is small
compared with the line-width of the cavity, the cavity is servoed at resonance with a realistic
servo control  and the variation in the
radiation pressure force is linearly dependent on the
displacement\cite{vij}. The radiation pressure effects have also been investigated in earlier  
literature \cite{mey,dor,deu,agu,mer}. Here, however, since we now have a reasonably good idea
about the instrumental parameters to be used in the large scale
 detectors, we expect that our analysis here will be important
to the experimentalists. We assume that the mirrors are hanging
`freely' (there is no active servo control) and the radiation pressure exerts
force on them which displaces them from  resonance. 
It is
very important for people involved in the experiments being built to have a quantitative idea of
the magnitude of these effects. 
For instance, it would be necessary to take into account these effects and modify the servo transfer functions to be used for the
control of the experiments.

The main result of this paper 
is to establish that the freely hanging mirrors continuously gain
energy and swing with ever increasing amplitude when subjected to radiation 
pressure force arising from the light field. This result seems to be in opposition with the results of Meystre et al., 
who found in \cite{mey} a mirror confinement due to the radiation pressure.
The reason for this difference of behaviour is the
`time delay' effect which is also examined in detail in this paper. The time delay is large for the long (kilometric) cavities,
 primarily studied here, while it can be neglected for short cavities such as the one studied by Meystre et al. \cite{mey}
and those used in most of labs. 
Closed form expressions 
have been given which facilitate in understanding the physics of the phenomenon.\\

The paper is organized as follows:\\
In section II, we set up the equation of motion of the free mirrors. 
We examine
the motion of the mirrors with the two forces $(i)$ the radiation
pressure force, $(ii)$ the force of gravity. In section III, we
numerically integrate the equations of motion using the so called
`Phase Space Method'. We present the results for the particular case,
when the mirrors are in the resonance positions and the laser is 
switched on. We observe that the amplitude of the motion of
the mirrors increases with time and energy is pumped into the
system. For very large times, when the amplitude of the system is also
very large, so that the mirrors cross several of the Fabry-Perot
resonances in one cycle of the pendulum, the motion approximates to
that of an anti-damped harmonic oscillator. We give also, for comparison,  the numerical results obtained for a short
cavity : no anti-damping is exhibited in this case, in agreement with earlier results \cite{mey}.
In section IV, we obtain analytically, under the quasi-static
approximation, the phase space trajectories of the motion of the
individual mirrors as well as the motion in the differential mode and the
common mode (the centre of mass mode). The analytical results match with the
numerical ones remarkably. 
In section V, we give a quantitative description of the phenomenon 
under the assumption that
the velocity of the mirror does not change very much on the time
scale of the storage time of the cavity. We find that the gain in
energy is due to a differential radiation pressure force arising from
the asymmetry, according as the mirrors are approaching each other or moving
away from each other. We obtain approximate analytical expressions
for the differential force, the time `delay' and then proceed to
compute the gain in energy per cycle when the mirrors encounter a single
resonance or cross several resonances. For large amplitudes,
when the mirrors cross several resonances, the system behaves like an
anti-damped harmonic oscillator. We can then associate an effective
negative $Q$-factor for the system. We show that the $Q$-factor depends
on the input power, 
the finesse of the cavity and the round trip time of the cavity. 
Finally in section VI, we study the behaviour of the system when it is initially in equilibrium and goes out of lock. 
This may happen when the servo loop
is suddenly opened.

\section{Optical and mechanical equations}
We consider only `free' mirrors meaning that no servo control
loop is used. The only forces acting on the mirrors are the radiation
pressure force and gravity which manifests itself as the restoring
force of the pendulum.
We consider a single cavity with mirrors $M1$ and $M2$ which are
suspended as shown in fig.1. The input beam $A$ enters the cavity from
mirror $M1$ and bounces back and forth between the two mirrors. After
several round trips, whose number is of the order of the finesse of the
cavity, the field builds up inside the cavity. The magnitude of the
field depends on the finesse of the cavity, the input power and the
detuning of the cavity. The field or the power produces the radiation
pressure force which pushes on the mirrors, driving them apart, thus
changing the distance between the two. This in turn changes the power
inside the cavity. 
 For instance, if the mirrors were hanging in a
position of resonance, the radiation pressure 
force drives the cavity out of resonance, reducing the radiation pressure 
force. The
mirrors start swinging with radiation pressure force adjusting to the
continuously varying length of the cavity. It is found that the
radiation pressure force does not adjust instantaneously to the new
length but {\em lags} behind the expected static force given by the Fabry-Perot
curve (the Airy function) by a time-lag comparable to the storage time of the cavity. The time-lag 
has been called `time delay' in earlier literature \cite{deu}.

  The slowly varying
amplitude of the field inside the cavity at time $t$, denoted by $B(t)$ satisfies 
the following equation,  
\begin{equation}
B(t) = t_{1}A\exp[ikx_{1}(t)] + R B(t-\tau)\exp[ikL(t)],
\end{equation} 
where, $x_{i}(t)$, $r_{i}$
and $t_{i}$, $i=1,2$ are the positions, reflectivities and
transmitivitties of the mirrors $M1$ and $M2$
respectively; $R=r_{1}r_{2}$, $k = 2\pi/\lambda$, where $\lambda$ is the
wavelength of the laser light, $\tau$ is the round trip time and 
\begin{equation} L(t) = 2x_{2}(t-\tau/2)-x_{1}(t)-x_{1}(t-\tau). 
\end{equation} 
In case of the VIRGO cavity,
the arm length $ L_{0}$ is  3 km, the round trip time $\tau\ = 2L_{0}/{c}
 \sim 2 \times 10^{-5}$ seconds, where $c$ is the speed of
light. This equation provides an iterative relation between the field amplitude
at time $t$ to the field amplitude at time $t-\tau$ and the positions of the
mirrors. We investigate the following two situations:
\begin{enumerate}
\item The mirrors are hanging in the positions of resonance and the
laser is switched on at the time $t=t_{0}$.
\item The mirrors are hanging in an equilibrium state with the
radiation pressure force balancing the restoring force of the
suspension.
\end{enumerate}
Situation 1 represents a possible experiment. When the servo-control is not operating, the mirrors will be freely
in motion. Then, when the laser is switched on, the radiation pressure
 will affect the motion of the mirrors. However, here, we mainly deal 
with the case (as given in 1), when the mirrors are initially at rest
and in the resonance position. However, many of our analytical formulae, for 
example, that of `time-delay' apply to more general situations.\\
Situation 2 describes the case when the interferometer is already in 
operation. Now if the servo-control loop is suddenly opened,
the system will tend to become unstable. This case is also investigated.

The equations of motion for the mirrors correspond to forced harmonic
oscillator with the forcing term arising from the radiation pressure
force. We first compute the radiation pressure forces on each mirror.
   For mirror $M1$, the radiation pressure force comprises of two 
terms, electric field due the input laser beam, $A$ and the intra-cavity 
field $B$ as shown in
fig.1. The radiation pressure force on $M1$ is,
\begin{equation}
F_{1}(t) = -\frac{2}{c}\bigl[R^{2}P(t-\tau)-r_{1}^{2}P_{0}\bigr], 
\end{equation}
where, $P(t)=|B(t)|^{2}$, $P_{0}=|A|^{2}$ and $c$ is
the speed of light. $r_{i}^{2}$ is the fraction of average 
number of photons reflected by $i$ th mirror for $i=1,2$ respectively.\\
For $M2$, the radiation
pressure force is given by
\begin{equation} F_{2}\Bigl(t-\frac{\tau}{2}\Bigr) = \frac{2 r_{2}^{2}}{c}P(t-\tau). \end{equation}
The equations of motion for the mirrors with the masses, natural
frequencies and damping constants, $m_{i}$, $\omega_{i}$, $\tau_{i}$;
$i=1,2$, respectively, are
\begin{equation} m_{i}\Bigl[{\ddot x}_{i}+\frac{2}{\tau_{i}}{\dot{x}_{i}}+\omega_{i}^{2}(x_{i} - 
x_{i0})\Bigr] = F_{i}(t), \end{equation} 
where $x_{i0}$ is the initial position of the mirror such that 
the separation between the mirrors before switching on the laser is 
$L_{0} = x_{20} - x_{10}$. The full system of equations to be evolved
in time are the
equations from (1) to (5) (non-linearly coupled equations). 
In section III, we first carry out the task
numerically and in later sections, after we have gained sufficient physical
insight into the problem,
we shall present the semi-analytical results. 

\section{The numerical solution}
For the numerical calculation, we consider the VIRGO parameters for the
suspension and the optical cavity. We assume $m_{1} = m_{2} = m \simeq
28$ kg., $\omega_{1} = \omega_{2} = \omega\simeq 3.75$ rad./sec. which corresponds to
a resonant frequency of about 0.6 Hz. The $Q$ factor for the suspension is typically of the order of
$10^{6}$. The optical parameters are $\tau\simeq 2\times 10^{-5}$ sec., the
wavelength of the carrier wave $\lambda\simeq1.064 \mu$m, $r_{1} \simeq 0.94$,
$r_{2}\simeq 1$. The wave number is $k=2\pi/\lambda$ and $R\simeq 0.94$. We examine
the behaviour of the system with the input power varying between 1 kW to
30 kW. Initial detectors will be operated at input powers $\sim$ 1 kW and
advanced detectors at powers of few hundred kW or even upto a MW.
 Note that these are the input powers for the main cavities after power recycling 
has been implemented.  
  
With the above values for the parameters, we
find that the instabilities set in, on the time-scales of few seconds
to few hundred seconds. Since the $Q$-factor of the pendulum suspension
 is so large, the damping in
the oscillations can be neglected for the numerical integrations carried
over the time intervals $\ll \frac{Q}{\omega} \sim 10^{6}$ seconds. Most of our
numerical integrations range from few seconds to at most few thousands of
seconds. Neglecting damping, the equations of motion for the mirrors
become,  \begin{equation} {\ddot x}_{i}+\omega^{2}(x_{i}-x_{i0}) = \frac{ F_{i}(t)}{m} =
f_{i}(t),\hspace{0.2in} i=1,2. \end{equation} The equations of motion of the
mirrors are non-linear differential equations and cannot be solved by
simple methods.   We integrate the equations by the so called
`Phase Space method' described below:\\ Let $\Delta$ be the time-step of
integration. The natural time step we assume is $\Delta = \tau$. We assume
that the forcing term is a constant during each time step. This
assumption is not unrealistic because the round trip time interval for 
the VIRGO cavity is of the order of $10^{-5}$ seconds; note that this approximation will 
hold {\it a fortiori} for more common (much shorter) cavities. The equations 
simplify enormously under this assumption. Thus we can integrate the equations 
exactly within this time interval.\\
The evolution of the equations goes as follows: 
\begin{equation}
x_{n+1} = x_{n}\cos\omega\Delta + p_{n}\sin\omega\Delta +\frac{f_{n}(1-\cos\omega\Delta)}{\omega^{2}}, 
\end{equation} 
\begin{equation} p_{n+1} = -x_{n}\sin\omega\Delta + p_{n}\cos\omega\Delta +
\frac{f_{n}\sin\omega\Delta}{\omega^{2}}, 
\end{equation} 
where we have dropped the indices 1,2 for
simplicity and $p = {\dot x}/\omega$. Here, $x$ represents the displacement from the mean position 
$x_0$. The index $n$ represents the value of the variable at the time
$n\Delta$, i.e. for example, $x_{n} = x(n\Delta)$.  
  The optical component has
the following iterative evolution:
 \begin{equation} f_{1n} = -\frac{2[R^{2}P_{n-1} - r_{1}^{2}P_{0}]}{mc}, \end{equation} 
\begin{equation} f_{2n} =\frac{ 2 r_{2}^{2}P_{n-1}}{mc}, \end{equation}
 \begin{equation} P_{n} = |B_{n}|^{2}, \end{equation}
 \begin{equation} B_{n} = t_{1}A\exp(ikx_{1n}) + R B_{n-1}\exp(ikL_{n}), \end{equation} 
\begin{equation} L_{n} = 2x_{2n} - x_{1n} - x_{1,n-1}. \end{equation}
This scheme solves the system of equations. Since it is the length of
the cavity that actually matters for this problem, we define the
variable, 
\begin{equation} \psi(t) = k[x_{2}(t)-x_{1}(t)] - k L_{0}, \end{equation} 
and present
the results in phase space plots of ${\dot \psi}/\omega$ vs $\psi$. $\psi(t)$ is 
called the differential mode.
  
We first consider the case when the mirrors are hanging in the
resonance position and the laser is switched on. As the power builds
up, the radiation pressure acts on the mirrors, driving them apart
resulting in the detuning of the cavity. This reduces the radiation pressure
force and the mirrors swing back. The motion is oscillatory and as
we note, the oscillations increase in amplitude. We employ two values
of input power namely, 1 kW (initial VIRGO) and 30 kW corresponding to
advanced detectors. The results are presented by the phase space
trajectories of the mirrors. We consider four variables for the
purpose, $x_{1}$, $x_{2}$, $\psi$ and  $\phi$. $\phi$, defined later in the text in 
eq. (24), section IV - B, is called the common mode.
For the range of powers considered 1 kW to 30 kW, the phase space curves obtained from
numerical simulations, for few tens of seconds are qualitatively
the same.\\
We make the following general observations about the features:
\begin{enumerate}
\item We observe from figs.4 and 5 that the radius of the phase space curve
increases with every cycle which indicates that the mirrors
continuously gain energy from the input laser beam implying that the system is
nonconservative. We also observe that the gain per cycle is not constant
but some sort of a periodic function of the radius of the
phase space diagram. We shall consider this phenomenon in detail in
section V. 
\item The (static) radiation pressure force peaks when the cavity is
in resonance and drops down to zero when it is out of resonance, (see fig.3).
 The full width at half maximum
(FWHM) of the radiation pressure force $F(\psi)$ is about $0.06$ rad, 
corresponding to the FWHM of the cavity resonance curve for a finesse of 50.
 Hence, for the initial stretch of the phase space trajectory,
the mirrors experience the radiation pressure force whereas during
the rest of the time, they only experience the restoring force (and the
force due to the input power for $M1$). The phase-space trajectory is
circular during the restoring force regime and is deformed away from the circularity 
when the mirrors encounter the appreciable amount of radiation 
pressure near resonance (see fig.4). 
\item Since the laser is beamed in the positive $x$-direction, there is 
an asymmetry about the origin. This shifts the centre 
of mass 
trajectory to the positive side of the $x$ axis (see fig.4). 
The period of oscillation of the common mode is twice that of the
period of the differential mode of the system (see fig.2). 
\item If we let the laser beam pump in energy for large amounts of
time, the amplitude also becomes large and the mirrors sweep across several
resonances. The phase space trajectory then tends to become more and
more circular and the motion approximates to that of a simple harmonic
motion. This feature can be observed in all the four modes. The
circularity of the trajectory implies that the motion is almost
`free'. The radiation pressure force has little effect because the
mirrors sweep too quickly across the resonances for it to affect their
motion. However, as we shall see that the steady gain in energy still
persists. Figure 5 depicts this phenomenon.
\item The amount of
energy imparted to mirror 2 by the laser beam after getting reflected, 
is more than to mirror 1. Thus M2 swings with larger amplitude as compared
to M1 (see fig.4) (the radiation force is larger on M2).
\item Let's turn now to the dynamics of a short cavity. The fig.6 shows the space of 
phase trajectory for the differential mode $\psi$, for a short cavity of length $L_0=30$ cm (instead of 3 km
for the other examples) and with the same optical and mechanical parameters and initial conditions as in fig.5 obtained
with the 3 km long cavity.
The round-trip time is here $\tau \simeq 2\times 10^{-9}$ s, and the simulation lasts for 500 s (corresponding to a huge number
of round trips in the cavity). At the contrary of the kilometric cavity, see fig.5, no energy gain is found.
This is in accord with the results of Meystre et al., who have previously developped a theory of radiation-pressure driven
cavities \cite{mey}. The physical reason for this difference of behaviour between a short (common) cavity and a kilometric one
is precisely due to the difference of length, as explained in the following sections. Time delay effects are thus of utmost importance
for very long cavities, while they can be neglected for usual ones (and so have been neglected in Meystre's theory \cite{mey}).

\end{enumerate}

\section{Phase space trajectories for the first cycle}
In this section, we obtain approximate closed form expressions for the
equations of the phase space trajectories for about a period of one cycle. 
For the VIRGO case, this turns out to be between one or two seconds. 
This analysis
could be useful in the context of the initial locking of the cavity. 
\subsection{The differential mode $\psi$}
The equations of motion of the individual mirrors, equations (3) to (5)
allow us to write the equation of motion of the system of mirrors in the
differential mode as, 
\begin{equation} 
{\ddot \psi} + \omega^{2}\psi = \frac{k}{m}[F_{2}(t)-F_{1}(t)]. 
\end{equation}    
In the quasi-static approximation,
\begin{equation} 
P(t-\tau) \simeq P(t) = \frac{P_{max}}{1+(\frac{2\mathcal{F}}{\pi})^{2}\sin^{2}\psi},
\end{equation} 
where $\mathcal{F} = \pi\sqrt{R}$$/(1-R)$ is the finesse of the cavity and 
\begin{equation} 
P_{max} = \frac{t_{1}^{2}P_{0}}{(1-R)^{2}}. 
\end{equation} 
Thus the equation for $\psi$ is, 
\begin{equation}
\frac{\ddot \psi}{\omega^{2}} + (\psi + \psi_{0}) =
\frac{F_{0}}{1+(\frac{2\mathcal{F}}{\pi})^{2}\sin^{2}\psi}, 
\end{equation} 
where
\begin{equation} 
F_{0} = \frac{2k(r_{2}^{2} + R^{2})}{m\omega^{2}c} P_{max},
\end{equation} 
and
\begin{equation} \psi_{0} = \frac{2k P_{0}r_{1}^{2}}{m\omega^{2} c}. 
\end{equation}
For the VIRGO cavity, we have the following numerical values
for the quantities:\\
 $\psi_{0} \simeq 0.88 {P_{0}}/{10 kW}$, $\mathcal{F} \simeq $ 50 and
$F_{0} \simeq 62.6{P_{0}}/{10 kW}$.\\
For $\mathcal{F} \gg$ 1, the term on the right hand side of equation (18) is 
non-zero only when $\psi \ll 1$. With the approximation $\sin \psi \sim \psi$, we 
can easily integrate equation (18) to get the phase space trajectory of the 
mirrors,
\begin{equation} \frac{{\dot \psi}^{2}}{\omega^{2}}+(\psi+\psi_{0})^{2} =
\frac{F_{0}\pi}{\mathcal{F}}\tan^{-1}(\frac{2\mathcal{F}}{\pi} \psi) + \psi_{0}^{2}.\end{equation}

For low powers like 1 kW, the approximation $\sin \psi \sim \psi$
works remarkably well and agrees with the numerically obtained
phase space trajectory.
In order to compare the analytical and numerical results, we compare the maximum value 
of $\psi$, namely $\psi_{max}$, of the trajectories for various input powers in fig.7.

\begin{enumerate}
\item
When $\psi$ is small that is near resonance, the equation of the
trajectory reduces to  
\begin{equation} 
\frac{\dot{\psi}^{2}}{\omega^{2}} = 2F_{0}\psi.
\end{equation} 
Thus the trajectory is parabolic in shape and passes through the origin. 

\item When $\psi \sim 1$, the
trajectory is a circle as expected since there is hardly any radiation
pressure force acting on the mirrors. The equation of the trajectory in
this regime is
\begin{equation} 
\frac{{\dot\psi}^{2}}{\omega^{2}} + (\psi + \psi_{0})^{2} = \frac{F_{0}\pi^{2}}{2\mathcal{F}} + \psi_{0}^{2}. 
\end{equation} 
\end{enumerate}

When the input power is very large $\sim 50$ kW, the trajectory does 
not maintain this simple shape. For example, when $P_{0}= 50$ kW, 
the trajectory is as shown
in fig.8.

 This is because the high powers make 
the mirrors cross several resonances in the first cycle itself 
and consequently the trajectory has more complex behaviour. We do not pursue 
this case here.

\subsection{The common mode $\phi$}
 In this sub-section, we study
the motion of the center of mass of the system of two mirrors. We
define the center of mass coordinate of the two mirrors as,  
\begin{equation} 
\phi = k (x_1 - x_{10}) + k (x_2 - x_{20}). 
\end{equation} 
The equation of motion of the
system in the center of mass coordinate is, 
\begin{equation} 
{\ddot \phi} + \omega^{2}\phi = \frac{k}{m}[F_{1}(t) + F_{2}(t)], 
\end{equation} 
where for $F_{1}(t)$ and $F_{2}(t)$ are given by equations (3) and (4).
 The equation of motion takes the form,
\begin{equation} 
\frac{\ddot \phi}{\omega^{2}} + (\phi - \psi_{0}) = \frac{F_{c}}{1+(\frac{2
\mathcal{F}}{\pi})^{2} \sin^{2}\psi}, 
\end{equation} 
where 
\begin{equation} 
F_{c} = \frac{2k(r_{2}^{2}-R^{2})t_{1}^{2}P_{0}}{mc\omega^{2}(1-R)
^{2}}.
\end{equation}
For VIRGO parameters, $F_{c} \simeq 4.0 \frac{P_{0}}{10 kW}$.  The above
equation is a second order differential 
equation and is coupled nontrivially to the $\psi$ mode. The strategy we adopt is to study the 
motion of the center of mass in different regimes; (1) near the resonance 
and (2) away from the resonance. The full trajectory is obtained by matching
the solution in the region
of the overlap.

\begin{enumerate}
\item For $0 \leq \psi \leq 0.5$; $\sin \psi \sim
\psi$, equation (26) becomes
\begin{equation}
\frac{\ddot \phi}{\omega^2} + (\phi -\psi_{0}) = \frac {F_{c}}{1 +
(\frac{2\mathcal{F}\psi}{\pi})^{2}}.
\end{equation} 
Further for $\psi \leq 3 \times 10^{-3}$, 
$(\frac{2\mathcal{F} \psi}{\pi})^{2} \ll 1$; we neglect 
$(\frac{2\mathcal{F}\psi}{\pi})^{2}$ as compared to 1 and obtain, 
\begin{equation} 
\frac{{\dot
\phi}^{2}}{\omega ^2} + \phi^{2} = 2 (F_{c} + \psi_{0}) \phi. 
\end{equation}
For the VIRGO cavity and input power 1 kW, $F_{c}\sim 0.4$. Neglecting
the quadratic term in $\phi$ as compared to the linear term, we see that
the motion of the center of mass describes a parabola for low values 
of $\psi$ and
$\phi$,  

\begin{equation} 
\frac{{\dot \phi}^{2}}{\omega^2} \simeq 0.97 \phi. 
\end{equation}

We compare the
slopes of the phase space diagrams in the differential mode as well as
the common mode. The phase space curve is more steeper in the common
 mode as compared to the differential mode. 
 
\item Away from resonance, $\psi \geq 0.5$, the power stored 
inside the cavity is almost zero. 
The equation of motion of the 
center of mass is given by, 
\begin{equation} 
{\ddot \phi} + \phi = \psi_{0}.  
\end{equation} 

The solution for the initial conditions $\phi = \phi_{0}, {\dot \phi} ={\dot \phi}_{0}$ is, 
\begin{equation} 
\frac{{\dot \phi}^{2}}{\omega^{2}} + (\phi -
\psi_{0})^{2} = \frac{{\dot \phi_{0}}^{2}}{\omega^{2}} + (\phi_{0} -
\psi_{0})^{2}. 
\end{equation}
\end{enumerate}

This solution must be matched to the solution in case 1. Suppose we match the solution at $\phi_0 = 0.5$, from 
equation (29) we have $\dot \phi /\omega \sim 0.7$. This gives the approximate solution. In general, the solution in this region is,
 \begin{equation} 
\frac{{\dot \phi}^{2}}{\omega^{2}} + (\phi -
\psi_{0})^{2} =  2 (F_{c} + \psi_{0}) \phi_0   + (\phi_{0} -
\psi_{0})^{2} - \phi_{0}^{2}.
\end{equation} 

The equations (29) and (33) describe the full solution for this mode.

The phase space trajectory for the individual mirrors can be obtained from the motion of the mirrors in the differential and the
common modes.  
The trajectories of the $x_1$ and $x_2$ modes are shown in figures 2 and 4.

\section{Energy considerations and anti-damping}

In the previous section, we examined the motion of the mirrors under
the radiation pressure force. We noted that each time when the mirrors cross a resonance they experience an impulsive force. In this section, we analyze in detail as to how the system gains energy 
as each resonance is encountered. 
\subsection{Quasi-static approximation}
We start with the quasi-static approximation in which the system is
conservative, thus there is no net gain in energy. We then
phenomenologically introduce the `time delay', $\tau_{lag}$ which now
leads to gain in energy. We thus obtain $\tau_{lag}$ in terms of the
cavity parameters and this gives us an equation for an anti-damped
harmonic oscillator. The energy gain can be obtained as shown in the
following sections. Finally, for large times, the gain can be expressed
through a negative $Q$ of a harmonic oscillator. 
  In the quasi-static approximation, we assume that the mirrors are moving 
`slowly' that is the intra-cavity power has time to adjust itself to the slowly 
changing positions of mirrors, i.e.  $\tau \dot{\psi} \ll 1-R$. Thus neglecting
$\tau$ from the equation of motion we obtain,

\begin{equation} 
\frac{\ddot \psi}{\omega^{2}} + (\psi + \psi_{0}) = F_{s}(\psi), 
\end{equation} 

where $\psi_{0}$ as given in the previous section gives the constant
displacement due to the constant input power $P_{0}$ from the laser
beam and $F_{s}(\psi)$
is the radiation pressure force in the static case is given by,  
\begin{equation} F_{s}(\psi) = \frac{F_{0}}{1 + (\frac{2\mathcal{F}}{\pi})^{2}\sin^{2}\psi}, \end{equation}
where $F_0$ is given in equation (19). 
The dimensionless energy of the system is an integral of motion and 
is obtained as   
\begin{equation} E = \frac{1}{2} \frac{{\dot \psi}^{2}}{\omega^{2}} + V_{sus}(\psi) + 
V_{0}(\psi) + V_{rad}(\psi), \end{equation} 
where the potentials are given by, 
\begin{equation} V_{sus}(\psi) = \frac{1}{2} \psi^{2}, \end{equation} 
\begin{equation} V_{0}(\psi) = \psi_{0} \psi, \end{equation} 
\begin{equation} V_{rad(\psi)} = -\int_{0}^{\psi} F_{s}(\psi)d\psi. \end{equation}
The system is conservative in this approximation. These results have been
obtained in the earlier literature \cite{deu}.  
  
However, we note from the results of numerical simulations that 
the system indeed gains energy with every cycle of oscillation. Moreover, 
this gain occurs when the radiation pressure force is appreciable i.e. 
when the system is near resonance. Most importantly, we observe from the
numerical simulations that radiation pressure force lags behind its
quasi-static value by a `time-lag' which we denote by $\tau_{lag}$. 
We
find that in the case of VIRGO, $\tau_{lag}$ varies from about $16\tau$ to
$30\tau$ as one climbs up the resonance curve from half its maximum to
the maximum. $\tau_{lag}$ is of the order of the storage time of the 
cavity. With this in mind we write the force $F(t) \sim
F_{s}(t-\tau_{lag})$ and obtain the 
following equation of motion,
\begin{equation} \frac{{\ddot \psi}}{\omega^{2}} + (\psi + \psi_{0}) = F_{s}(\psi(t-\tau_{lag})). \end{equation} 
Taylor expanding the forcing term to the first order we obtain,
\begin{equation} \frac{\ddot \psi}{\omega^{2}} + \tau_{lag} \frac{dF_{s}}{d\psi} {\dot
\psi} + (\psi + \psi_{0}) = F_{s}(\psi), \end{equation}
as the equation of motion for the system. 
The ${\dot \psi}$ term in
equation (41) is responsible to the gain/loss of energy of the system. This
solely depends on the sign of $\frac{dF_{s}}{d\psi}$ since $\tau_{lag}$ is always positive. If we start the system from resonance at $\psi = 0$,
$\psi$ starts increasing slowly and in the region, $\frac{dF_{s}}{d\psi}< 0$ and the system experiences more force than what it would have been in the
quasi-static case. This excess force is manifested in an excess amount of
energy $\Delta E$. In general, the energy gained/lost over a certain amount of time
is given by,
\begin{equation} \Delta E = -\int_{\psi_1}^{\psi_2}
\tau_{lag}(\psi) \frac{d F_{s}}{d\psi} {\dot \psi} d\psi = -\int _{t_{1}}^{t_{2}} \tau_{lag}\frac{d F_{s}}{d\psi} {\dot \psi}^{2}dt. \end{equation} 
It is clear from equation (42) that when $\frac{dF_{s}}{d\psi} < 0$, the
energy gain $\Delta E > 0$ and vice-versa. But since the radiation
pressure force always gives a kick in the positive -$\psi$ direction,the
$|{\dot \psi}|$
is always larger when $\frac{dF_{s}}{d\psi} < 0$. In other words, the amount of energy
gained by the system is more than the amount of energy lost. Hence, there is always a
net gain in energy when the system crosses a resonance.    We need to compute this
extra force $\Delta F$ which gives rise to the gain in the energy to the first order in
$\tau_{lag}$. We obtain,
 \begin{equation} \Delta F = -\frac{dF_{s}}{d\psi} {\dot \psi} \tau_{lag}. \end{equation} We note that $\Delta F$
depends on the phase velocity and $\frac{dF_{s}}{d\psi}$. Since $\frac{dF_s}{d\psi} \simeq 0$
when the system is away from resonance, $\Delta F$ comes into play only in the 
region of resonance. 
Another way of describing the fundamental asymmetry is to say that for the moving
mirror, the wavelength of the laser light is modified by the Doppler effect.
When the mirrors are approaching each other, the apparent frequency of the light for the cavity is
 increased,
or equivalently, the line-width of the cavity seen from the laboratory
frame is narrower and vice versa
when the mirrors move away from one another, the line-width is seen to be
broader. The consequence is
that the braking force (when the mirrors move against the light) acts for a
shorter time than the accelerating force, when the mirrors are moving away from
each other. Over one cycle the energy difference is positive and there is a continuous increase of mechanical energy which comes from the laser.

\subsection{The computation of the time-delay} 
In this section, we obtain a
closed form expression for the `time-delay', $\tau_{lag}$ under the
approximation that the relative velocity of the mirrors does not
change much during the storage time of the cavity. This is observed in
the numerical simulations and hence the approximation may be justified. 
$\tau_{lag}$ is
the most important quantity for computing the rate of gain in
energy.
  The intra-cavity radiation field at $n$-th time instant
$B_{n}$ is related to the intra-cavity field at $n+1$-th time instant
via an iterative relation, 
\begin{equation} B_{n+1} = t_{1}A +
R\exp(2i\psi_{n})B_{n}, \end{equation} where, $B_{n} = B(n\tau)$, $\psi_{n} = \psi(n\tau)$.
 We compute the field $B_n$ from initial time $t_0$, which
we take to be zero. We take the time step for the iteration to be the
round trip time $\tau$. In order to compute the intra-cavity field, it
is necessary to know the temporal behaviour of $\psi$. For the static
case, when the mirrors are stationary, we get the equilibrium field
$B_{s}$.  When the mirrors are moving, the approximation now comes into play,
namely, we assume that
$\dot{\psi}$ is constant over the storage time of the cavity.\\ 
The Taylor expansion
of $\psi_{k}$ to the first order around $t = 0$ is,
\begin{equation} \psi_{k} = \psi_{0} + k\tau{\dot \psi}_{0}, 
\end{equation}
where $\psi_0 = \psi(t=0).$
Iterating equation (44) $n$ times starting from $t = 0$ when $B = B_{0}$, 
we obtain, 
\begin{equation} 
B_{n} = t_{1}A\sum_{m=0}^{n-1}R^{m} e^{2i\sum_{k=n-m}^{n-1}
\psi_{k}} + R^{n}e^{2i\sum_{k=0}^{n-1}\psi_{k}}B_{0}. 
\end{equation}  
For large $n$, the last
term  in equation (46) tends to zero and the electric field
amplitude at the $n$-th iteration is given by, 
\begin{equation} B_{n} = t_{1}A
\sum_{m=0}^{n-1}R^{m} e^{2i\sum_{k=n-m}^{n-1}\psi_{k}}. 
\end{equation}

\begin{itemize} 
\item If the relative separation between the mirrors is constant in
time, $\psi_{k} = \psi_{0}$, we retrieve the equilibrium field $B_s$: 
\begin{equation} 
B_{s} = t_{1}A\sum_{m=0}^{\infty}R^{m}e^{2im\psi_{0}} = \frac{t_{1}A}{1-R e^{2i\psi_{0}}}.
\end{equation}
\item Turning now to the case in hand, when $\psi_{k}$ varies linearly with time, 
the sum in the exponential for large $n$ is approximately given by
\begin{equation} 
\sum_{k=n-m}^{n-1} \psi_{k} \sim m\psi_{n}-\frac{1}{2}m^{2}\tau\dot{\psi_{0}}.
\end{equation}
We combine equations (47) and (49) to obtain,
\begin{equation} 
B_{n} = t_{1}A
\sum_{m=0}^{n-1}R^{m}e^{2im\psi_n}e^{-im^{2}\tau\dot{\psi_{0}}}.
\end{equation} 
If $\dot{\psi}$ is small then the exponent
can be linearized. Then it is possible to express the field as the sum
of the static term $B_s$ and
the remaining part $\Delta B$ which corresponds to the time lag.\\
We write   
\begin{equation} 
B_{n} = B_{s} + \Delta B, 
\end{equation} where
\begin{equation}
\Delta B = -i\tau t_{1}A\dot{\psi_{0}} \sum_{m=0}^{n-1} R^{m} e^{2im\psi_{n}}m^{2}.
\end{equation}
Summing the arithmetico-geometric series \cite{gra} yields,
\begin{equation}
\Delta B = -2it_{1}A \tau \frac {\dot{\psi} R^{2}e^{-4i\psi}}{(1-Re^{-2i\psi})^{3}}.
\end{equation}
The corresponding power $\Delta P$ is given by
\begin{equation}
\Delta P = 2 Re(B_{n} \Delta B_{n}^{*})\sim \frac{16t_{1}^{2}|A|^{2}
\tau R^{2} \psi \dot{\psi}
(1-R)}{[(1-R)^{2} + 4R\psi^{2}]^{3}}.
\end{equation}
Refering to equation (19), we get the expression for the dimensionless
extra force,
\begin{equation} 
\Delta F = \frac{2k(r_{2}^{2} + R^{2})}{m \omega^{2}c} \Delta P.
\end{equation}
 The above expression of power
is for values of $\psi$ near resonance when $\psi \ll
1$. When $\psi\sim n\pi$, the same expression can be replaced by the $\psi
- n\pi$ under similar approximations. The effective `time delay' $\tau_{lag}$ is now obtained from (43), (54) and (55).
\begin{equation} 
\tau_{lag} = \frac{2\tau R}{(1-R)(1+(\frac{2\mathcal{F}}{\pi})^{2}\psi^{2})}. 
\end{equation} 
\end{itemize}
We note from the equation (56) that the effective time lag is maximum
at the resonance and starts decreasing as one goes away from the
resonance value.  For VIRGO parameters, we compute the value of the
$\tau_{lag}$. At FWHM, i.e. $\psi \sim 0.03$, $\tau_{lag}\sim 16\tau$,
going up to $\tau_{lag} \sim 30\tau$ as one approaches resonance.

\subsection{Energy gain near the resonance at $\psi=0$}
 In this section, we examine the motion of the mirrors for small 
amplitudes. Initially before the laser is switched on, the cavity 
is at resonance i.e. the two mirrors are separated by an integral 
multiple of $\pi$ in phase ($\frac{kL_{0}}{\pi}$ is an integer). 
The system starts from $\psi = 0$ and the motion is allowed to 
evolve with time. We restrict the amplitude to $|\psi| \leq \pi$ and 
study the system in this regime. Our goal is to compute the net
gain in energy, 
$\Delta E_{cycle}$ during one cycle of oscillation.
We have,
 \begin{equation} 
\Delta E_{cycle} =
2\int_{0}^{\psi_{1}} \Delta F(\psi) d\psi, 
\end{equation} where  
\begin{equation}  
\Delta F(\psi) =
\frac{16\tau R^{2} F_{0} \psi {\dot
\psi}}{(1-R)^{3}(1+(\frac{2\mathcal{F}\psi}{\pi})^{2})^{3}}. 
\end{equation}
We note that $\frac{d F_{s}}{d\psi} \leq 0 $ in this regime. We set an 
arbitrary  cut-off $\psi_{1}$ (when $\Delta F \ll 1$) as the system moves 
away from the resonance. The factor of two is because
the force is encountered twice during the cycle. \\ 
The $\dot{\psi}$ is obtained from the energy balance equation,
\begin{equation} 
\frac{1}{2}\frac{ {\dot \psi}^{2}}{\omega^{2}} =
\int_{0}^{\psi} F_{s} (\psi) d\psi \sim \frac{\pi}{2\mathcal{F}} F_{0}
\tan^{-1}(\frac{2\mathcal{F}\psi}{\pi}). 
\end{equation} 
Thus from (57) and (58) we get,
\begin{equation} 
\Delta E_{cycle} = 8\omega \tau \sqrt{\frac{R\mathcal{F}}{\pi}}
F_{0}^{3/2}I(\phi_{1}), 
\end{equation} 
  where, 
\begin{equation} I
(\phi_{1}) =
\int_{0}^{\phi_{1}} \frac{\phi(\tan^{-1}\phi)^{1/2}}{(1+\phi^{2})^{3}}
d\phi,\end{equation} 
and $\phi_{1} = 2{\mathcal{F}}{\psi_{1}}/{\pi}$. The cut-off $\phi_{1}$ 
should be away from the resonance and we find for VIRGO, $\phi_{1} \sim 5$ is an acceptable value.

 The numerical and the analytical results are compared 
in fig.9 by plotting the gain in energy per cycle for various input powers.

\subsection{Energy gain for large amplitudes}
As the mirrors gain energy, they swing with ever increasing amplitude
and sweep over several resonance peaks of the Fabry-Perot cavity. The system
gains energy at every resonance peak and thus the energy gained per cycle is the
sum of the energy gained at each resonance encountered. The peaks are 
encountered at $\psi = n\pi$, with $n_{max}\geq n\geq n_{min}$,
$n_{min}<0$ and $n_{max}>0$. The total number of resonances
encountered by the mirrors is $n_{max}+|n_{min}|+1$. Also it is
observed that $n_{max} \geq |n_{min}|$ due to the built-in asymmetry
arising due the laser power pumped in the positive $x$-direction (see fig.3).\\
Let $\Delta E_{n}$ be the energy gain at the $n$-th resonance, then the total
energy gained per cycle $\Delta E_{cycle}$ is given by,
\begin{equation}
\Delta E_{cycle} = \sum_{n_{min}}^{n_{max}} \Delta E_{n},
\end{equation}
where
\begin{equation} \Delta
E_{n} = \Delta E_{n+} + \Delta E_{n-}, \end{equation} and $\Delta E_{n-}$, $\Delta E_{n+}$ is the
energy
lost or gained respectively and  given by the following expressions,
\begin{equation} \Delta E_{n+} = \int_{n\pi}^{n\pi + \psi_{1}} \Delta F d\psi \hspace{0.1in}
> 0,\end{equation}
\begin{equation}\Delta E_{n-} = \int_{n\pi-\psi_{1}}^{n\pi} \Delta F d\psi \hspace{0.1in} < 0.\end{equation} 
Combining
equations (57,58,62 to 65) we get the energy gain at $n \pi$ as,
\begin{equation} \Delta E_{n} \simeq
 \frac{16\tau R^{2} F_{0}}{(1-R)^{3}}({\dot \psi}_{n+} - {\dot
 \psi}_{n-})\int_{0}^{\psi_{1}}
 \frac{\psi}{(1+(\frac{2\mathcal{F}\psi}{\pi})^{2})^{3}} d\psi, \end{equation} where
 $\dot{\psi}_{n-}$ and $\dot{\psi}_{n+}$ are the relative phase 
velocities of the mirrors for $\psi \leq n\pi$ and 
$\psi \geq n\pi$ respectively.
The energy gained at different resonance positions of the 
mirrors is different because the $\dot \psi$ is different at 
different resonances, $\dot \psi$ is maximum when $|n|$ is small and becomes small when $n$ approaches $n_{min}$ or $n_{max}$.
Thus the net energy gain per cycle is 
\begin{equation}
 \Delta E_{cycle} \simeq \frac{\tau R^{2} F_{0}}{(1-R)^{3}}
\bigl(\frac{\pi}{\mathcal{F}}\bigr)^{2}\sum_{n_{min}}^{n_{max}}({\dot \psi}_{n+} - {\dot
\psi}_{n-}). \end{equation}   The next task is to compute the $({\dot \psi}_{n+} - {\dot
\psi}_{n-})$, the increase in phase space velocity while 
crossing the resonance position as a function of $n$. For large values of $n$, the radiation pressure effect reduces
remarkably as is observed in fig.3. 
Hence we take the static force equation to
compute the increase in the phase space velocities. We 
integrate equation (41) neglecting the anti-damping term
obtaining the
change in kinetic energy as,  
\begin{equation} 
\frac{1}{2\omega^{2}}({\dot \psi}_{n+}^{2} - {\dot
\psi}_{n-}^{2}) = F_{0}\int_{-\infty}^{\infty}
\frac{1}{1+(\frac{2\mathcal{F}}{\pi})^{2}\psi^{2}}d\psi - 2\psi_{1}\psi_{0} -
2n\pi\psi_{1}. \end{equation} Equation (68) reduces to \begin{equation} \frac{({\dot
\psi}_{n+} - {\dot \psi}_{n-})}{\omega} =\frac{\frac{\pi^{2}F_{0}}{\mathcal{F}}
-4\psi_{1}(\psi_{0}+n\pi)}{2\dot\psi(n\pi)/\omega},
\end{equation}
where we have approximated $\dot \psi_{n+} + \dot \psi_{n-} \simeq 2 {\dot \psi}(n\pi).$\\
 To compute ${\dot \psi}(n\pi)/\omega$ analytically, we consider
the dimensionless
instantaneous energy as given by equation (36),

\begin{equation} {E} = \frac{1}{2} \frac{{\dot
\psi}^{2}}{\omega^{2}} + \frac{1}{2} (\psi^{2} + 2\psi \psi_{0}) -
\frac{2k(r_{2}^{2} + R^{2}) t_{1}^{2} P_{0}}{m\omega^{2}c} J(\psi), 
\end{equation} 

where
$J(\psi) = \int_{0}^{\psi} \frac{d\psi}{(1-R)^{2} + 4R\sin^{2}\psi}$.  We can approximate the
integral for large motions as $J(\psi) \sim \frac{n\pi}{1-R^{2}} \sim \frac{\psi}{1-R^{2}}$
for $n$ crossings of the resonances. Assuming $r_{2}=1$ and $r_{1}=R$, we have,
 \begin{equation} \frac{{\dot \psi}^{2}}{\omega^{2}} + (\psi -
\psi_{c})^{2} = (2E +\psi_{c}^{2}), \end{equation} 
where $\psi_{c} = \frac{2kP_{0}}{m\omega^{2} c}\sim \frac{P_{0}}{10 kW}$ (for VIRGO parameters). The phase space
trajectory is a circle centred around $\psi_{c}$ with radius 
\begin{equation} \rho = \sqrt{2E + \psi_{c}^{2}}.\end{equation}
 Equation (71) gives
 $\frac{\dot \psi(n\pi)}{\omega}$ as 
 \begin{equation} \frac{{\dot \psi}(n\pi)}{\omega} = \sqrt{ \rho^{2} - (n\pi
-\psi_{c})^{2}}. \end{equation} We rewrite the gain in energy per cycle as,
  \begin{equation}  \Delta E _{cycle} = \frac{\tau R^{2} F_{0}
\omega}{(1-R)^{3}}\bigl(\frac{\pi}{\mathcal{F}}\bigr)^{2} S(n_{max},n_{min}),\end{equation}
where $n_{max}$ is the greatest integer not greater then 
$(\psi_{c} + \rho)/\pi$ and $n_{min}$ is the smallest 
integer not smaller then $(\psi_{c} - \rho)/\pi$. We have, 
\begin{equation}
S(n_{max},n_{min}) = \sum_{n=n_{min}}^{n_{max}} g(n\pi),\end{equation}
where
\begin{equation}g(n\pi) = 
\frac{\frac{\pi^{2}F_{0}}{2\mathcal{F}}-2\psi_{1}(\psi_{0}+n\pi)}{\sqrt{\rho^{2} -
(n\pi -\psi_ {c})^{2}}}.\end{equation}  For large motion of the mirrors, since
the mirrors cross many resonances, we may replace the sum by an
 integral over $n$. Changing over to the variable $M = n\pi - 
\psi_c$, in which the system appears more symmetric about the 
origin, we have,
\begin{equation} S = \int_{M_{min}}^{M_{max}} \frac{\alpha - \beta M}{(\rho^2 - M^2)^
{1/2}} dM \end{equation}
where $\alpha = \frac{\pi F_0}{2\mathcal{F}} - \frac{2\psi_1}{\pi}(\psi_0 +
\psi_c)$ and $\beta = \frac{2\psi_1}{\pi}$. $M_{max}$ and $M_{min}$
correspond to $n_{max}$ and $n_{min}$ respectively.\\
We note that $\rho$ satisfies the following inequalities,
 \begin{equation} M_{max} \leq \rho \leq M_{max} +1, \hspace{0.2in} |M_{min}| \leq
\rho \leq |M_{min}|+1.\end{equation}
We observe that when $n$ is large, the difference between $|M_{min}|$ 
and $M_{max}$ is 
small, of the order of 1 or 2 times $\pi$. We denote the difference, by
$\delta M$ where,
\begin{equation} \delta M = M_{max} - |M_{min}| = M_{max} + M_{min}.\end{equation}
The integral in equation (77) splits into three parts $S = S_1 -
S_2 - S_3$, where
\begin{equation} S_1 = 2\alpha \sin^{-1} M_{max}/\rho,\end{equation}
\begin{equation} S_2 \simeq \delta M \frac{2\alpha}{(\rho^2 - M_{max}^{2})^{1/2}},\end{equation}
\begin{equation} S_3 \simeq \beta \frac{M_{max} \delta M}{(\rho^2 - M_{max}^{2})^
{1/2}}.\end{equation}  
The dominant term is $S_1$ which gives the general behaviour
and shape of the curve shown in fig.10. While $S_2$ produces a small kink 
in the curve, the effect of
$S_3$ can essentially be ignored. In case of the VIRGO cavity and $P_0 = 30$ kW
and $\rho \simeq 20$, $S_1 \simeq 16.7$, $S_2 \simeq 1.75$ and $S_3
\simeq 0.26$. For the initial VIRGO detector, $P_0 = 1$ kW, 
and $\rho \simeq 4.0$, $S_1 \simeq 0.56$ while $S_2, S_3 \simeq 0$.

Considering only $S_1$, the energy gain per cycle is approximately given
by,
 \begin{equation} \Delta E_{cycle} = \frac {2 \Delta E_{max}}{\pi} \sin^{-1} \frac{M_{max}}{\rho}, \hspace{0.2in}
M_{max} \leq \rho \leq M_{max}+1, \end{equation} where, \begin{equation} \Delta E_{max} = \frac {\tau R^{2} F_{0} \omega
\pi^{3}}{(1-R)^{3}{\mathcal{F}}^{2}} \Bigl(\frac{\pi F_{0}}{2\mathcal{F}} 
-\frac{2\psi_{1}}{\pi}(\psi_{0} + \psi_{c})\Bigr).\end{equation}
We observe the following features in the profile of $\Delta E_{cycle}$:
\begin{itemize}
\item 
At the resonance position
, $\rho = M_{max}$ the energy gain is maximum and equal to $\Delta E_{max}$.
For the
VIRGO cavity specifications and input power of 30 kW, $\Delta E_{max}\simeq
3.8$. When $P_0 = 1$ kW, $\Delta E_{max}\simeq 4.2 \times 10^{-2}$. 
\item Equation (83) shows that the energy gain per cycle is a
decreasing function of $\rho$. The energy gain decreases till the next
resonance is crossed, where it suddenly increases to $\Delta E_{max}$.
As the amplitude increases, $\rho$ increases from $M_{max}$ to
$M_{max} + 1$, the mirrors sweep across the resonances a little
faster which deprives them from gaining the full energy 
$\Delta E_{max}$. $\Delta E_{cycle}$ therefore reduces from $\Delta E_{max}$ to
$\Delta E_{min}$, where 
\begin{equation} \Delta E_{min} = \frac{2\Delta E_{max}}{\pi} \sin^{-1}\frac{M_{max}}{M_{max} +1}.\end{equation}
\item The energy profile in this range of $\rho$ i.e.$M_{max}$
to $M_{max} +1$ can be approximately given by,
\begin{equation} (\Delta E_{max} - \Delta E_{cycle})^{2} = \frac{8(\Delta E_{max})^{2}}{\pi^{2} M_{max}}(\rho - 
M_{max}).\end{equation}

We note that the minimum value is,
\begin{equation}\Delta E_{min} \sim \Delta E_{max} \Bigl( 1 - \Bigl(\frac{8}{\pi^2 M_{max}}\Bigr)^{1/2}\Bigr),\end{equation}
and it tends to the maximum value $\Delta E_{max}$ as $M_{max} $ becomes very large.
\item In fig.10, it is seen that there is a kink after $\Delta E_{min}$ is reached . The kink occurs because $M_{min}$ reduces by 1 when the mirrors cross
yet another resonance on the negative side.
This is accounted for by the second term.
\end{itemize} 
Since the $E \sim \rho^{2}$, the rate of increase of $\rho$ per 
cycle, denoted by $\Delta \rho$ is given by $\Delta \rho = \frac{\Delta E}{ \rho}$. 

\subsection{The negative Q-factor}
In the previous section, we have seen that for large values of $\rho$,
the amount of energy gained by the system of two mirrors per
cycle is a constant and tends to $\Delta E_{max}$. The system is an anti-damped harmonic oscillator 
which gains on an average, constant amount of energy per cycle. We 
may therefore associate a {\em negative} quality factor $-Q$, where $Q > 0$, with the system which 
describes the anti-damping. In this section, we endeavor to study the 
behaviour of $Q$.\\ 
The equation of motion of an anti-damped harmonic
oscillator is given by,
 \begin{equation} \frac{\ddot {\chi}}{\omega^{2}} - \frac { \dot{\chi}}{\omega Q} +
\chi = 0, \end{equation} where $Q$ is constant.
The solution for $\chi$ is of the form 
\begin{equation}\chi \sim \chi_0 e^{\omega t/2 Q} e^{\pm i\omega t}.\end{equation}
where
$\chi = \psi - \psi_c$. The amplitude is $\rho \sim \chi_{0} e^{\omega t/2Q}$ when $\rho$ is 
sufficiently large so that the phase space trajectory is approximately circular and can be compared 
to a simple harmonic oscillator.
Since a constant amount of energy is gained per cycle, the $Q$ will be a function of time. 
However, we can still describe the system by an average $Q$ taken over a cycle, which we 
denote by $<Q>_{cycle}$, since the change in $Q$ during one cycle is small. 
Moreover, we assume $\frac{1}{\omega <Q>_{cycle}} \frac {d <Q>_{cycle}}{dt}
\ll 1$ i.e. the fractional variation of $<Q>_{cycle}$ over a period of a cycle can be ignored.
Thus we obtain a W.K.B. solution for $\rho (t)$ as,
\begin{equation} \rho (t) = \rho (t_0) \exp\Bigl[\frac{1}{2} \int_{t_0}^{t} 
\frac{\omega dt}{<Q>_{cycle}}\Bigr], \end{equation}
where $t_0$ is some fixed but arbitrary initial time instant and $\rho (t_0)$
is the radius of the circular phase-space trajectory at $t_0$. We have also assumed 
that $<Q>_{cycle} \gg 1$.  
From
equations (72) and (90) the energy of the harmonic oscillator is given by,
\begin{equation}
E(t) - E(t_0) = \frac{1}{2} \rho^{2}(t_0) \Bigl[\exp \bigl(\int_{t_0}^{t}
\frac{\omega dt}{<Q>_{cycle}}\bigr)-1\Bigr].
\end{equation}
Since the energy gain is a constant and during a cycle equal to $\Delta E_{max}$,
we can equate $\frac{2\pi}{\omega}\frac{dE}{dt}$ to $\Delta E_{max}$ to obtain,
\begin{equation} E(t) - E(t_0) = \frac{\omega \Delta E_{max}}{2 \pi} (t - t_0).
\end{equation}
The time evolution of $\rho$ is obtained, which yields, 
\begin{equation}
\rho(t) = \rho (t_0) \big [ 1 + \frac{\Delta E_{max}}{\pi \rho^{2}_0} \omega(t-t_0) \big ]^{1/2}.
\end{equation}
Equating the logarithmic derivatives of (90) and (93) we obtain,
\begin{equation}
<Q>_{cycle}(t) = <Q>_{cycle}(t_0) + \omega (t-t_0), \end{equation}
where $<Q>_{cycle}(t_0) = \frac{\pi \rho^{2}(t_0)}{\Delta E_{max}}$.
We observe that both the energy and the $<Q>_{cycle}$ increase linearly with time while 
the amplitude $\rho$ increases as $t^{1/2}$. $<Q>_{cycle}(t)$
depends through $\Delta E_{max}$ on the input power, finesse and the round trip time.
 
For $P_0 = 30$ kW of input power, the trajectory more or less obtains a circular shape when $\rho (t_0) \sim 15$. For the VIRGO parameters,
\begin{equation} \Delta E_{max} \sim 0.42 (\frac{P_{0}}{10 kW})^{2} \sim 3.79.\end{equation}
Thus $<Q>_{cycle}(t_0) \sim 186$ and so 
\begin{equation} <Q>_{cycle}(t) \sim 186 + \omega (t-t_0).\end{equation}

Further, if we also consider the effect of the damping of the suspension then the
limit cycle will be approached when $<Q>_{cycle} \sim Q_{sus} \sim 10^{6}$ for VIRGO.
 $<Q>_{cycle}$ will attain this value after $\omega t \sim 10^{6}$ which corresponds to
little more than 3 days. The corresponding amplitude is given by,
\begin{equation}\rho = \rho(t_0) \bigl[1 + 
\frac {\omega (t-t_0)}{<Q>_{cycle}(t_0)}\bigr]^{1/2} \sim 1100.\end{equation}

This simple analysis will have to be modified when the limit cycle is almost reached, 
that is when, $<Q>_{cycle} \sim  Q_{sus}$. Here, however our goal was to estimate the 
time it takes to reach this stage. The above analysis is then adequate for the purpose.
However, when the servo operates or otherwise such a situation is unlikely to 
arise because other effects such as mirror tilting etc. will be important long 
before and then the dynamics will be completely different.

\section{The equilibrium case}

Lastly, we examine the case when the mirrors are in equilibrium under
the radiation pressure force and the restoring force. We have to
determine whether the equilibrium is stable or unstable. To this end,
we perturb the equation (41) about a given equilibrium
point. Writing $\psi = \psi_{eq} + \delta \psi$ and linearizing, we
get,
\begin{equation} \delta \ddot \psi - {2 \over \tau_{eq}} \delta \dot \psi +
\Omega^{2} \delta \psi = 0, \end{equation} where, \begin{equation} \tau_{eq} =
-\bigl(\frac{\pi}{2\mathcal{F}}\bigr)^{3}
\frac{(1+(\frac{2{\mathcal{F}}\psi_{eq}}{\pi})^{2})^{3}}{F_{0}\psi_{eq}\tau
\sqrt{R}}, \end{equation} and \begin{equation} \Omega_{eq}^2 = 2\bigl(\frac{2\mathcal{F}}{\pi}\bigr)^{2}
\frac{\psi_{eq}F_{0}}{(1+(\frac{2{\mathcal{F}}{\psi_{eq}}}{\pi})^{2})^{2}}. \end{equation} If
$\psi_{eq} \leq 0$, $\Omega_{eq}^2 \leq 0$, the sign of $\tau_{eq}$ does not
matter and the instability grows exponentially. If on the other hand
$\psi_{eq} > 0$, then although $\Omega_{eq}^2 > 0$, $\tau_{eq} > 0$ and
this leads to gradually growing oscillations until the pendulum tips
over the maximum. We plot the phase space trajectory for the input
power of 1 kW in fig.11. We conclude from this that the
cavity is {\it always} unstable when radiation pressure forces
act. The time-delay plays a crucial role in making the system
unstable.    The negative Q-factor for the motion of
the mirrors near the equilibrium position, is given by, 
\begin{equation} Q = - \frac{1}{2}\Omega_{eq}\tau_{eq}.\end{equation} 

For an input power of 30 kW, with VIRGO cavity
parameters, the equilibrium position of the mirrors near the resonance
at zero is $\psi_{eq}\simeq 0.25$. The corresponding values of the other
quantities are 
$\Omega_{eq}\simeq 4.75$ and $\tau_{eq}\simeq 10^{4}$ seconds. The negative Q-factor, 
$Q \simeq -2.3\times 10^{4}$. Whereas for initial VIRGO, 
$P_0 \simeq 1$ kW, $\psi_{eq} \simeq 0.16$ 
thus $\Omega_{eq}\simeq 1.67$, $\tau_{eq}\simeq 3.1 \times 10^{4}$ seconds, 
the negative Q-factor, $Q \simeq -2.6\times 10^{4}$.

\section{Conclusion}
We have analysed the effect of radiation pressure on the freely hanging 
mirrors (no servo loop) suspended in the laser interferometric optical 
cavities. After numerically evolving the full set of equations of motion
with respect to time, we find that the amplitude of the mirror oscillations
continuously increases as time progresses. We introduce the `time delay',
that is the time taken for field to adjust to the motion of the mirrors, in a 
phenomenological way to explain the observed gain. We conclude that the 
gain in energy
is due to the differential radiation pressure force arising from the 
asymmetry depending upon the motion of the mirrors. From another viewpoint,
we can also
explain the gain in energy qualitatively by the Doppler effect.
With respect to the mirror, 
the frequency of the incoming laser beam is higher as compared to that of the
outgoing laser beam due to the Doppler effect. The deficit of the energy of
the laser beam after getting reflected from the mirror can be looked upon
as the energy gained by the mirrors.
The values of the energy gain per cycle are computed analytically under the
reasonable assumption that the mirrors are not accelerated within the time scale of the
storage time of the cavity. For VIRGO parameters, the analytical values
agree remarkably with the numerical values. The interesting point to note is that the motion of the mirrors 
approaches that of an anti-damped
harmonic oscillator with a constant gain in energy as time progresses which
implies that the mirrors move too quickly to get affected by the radiation pressure force. The negative Q-factor of the anti-damped oscillator depends on the
input power, the finesse and the round trip time of the cavity and increases 
linearly as a function of time. The analysis will have to be modified when the
negative Q-factor becomes of the order of the damping Q-factor of the
suspension fibre, if such a case can arise.\\

In this paper, we have shown that the radiation pressure force makes the 
freely hanging mirror unstable {\em for all values of the input power and
irrespective of the initial conditions.}\\
The above analysis is relevant in the event, when the interferometer is in operation
and if the servo loop is suddenly opened. Then the motion of the hanging
mirrors can be deduced from the
above analysis. This analysis will be helpful in designing a servo-control which can prevent
this instability. In a previous work \cite{vij}, the servo-control was included in the linear regime of 
the Fabry-Perot curve assuming the transfer function
for the servo given by Caron et al. \cite{car}. Their work sets the stage for analysing the 
system in the non-linear regime as well, but it is then
needed to know how servo-control operates in the full regime.   

\section{Acknowledgement}
The authors thank the IFCPAR (project no. 1010 - 1) under which
a substantial part of this work was carried out. One of us (SVD) thanks the 
PPARC, U.K. for financial support, where this work was finished.

\onecolumn

\begin{figure}
\label{fig1}
{
\centerline{\psfig{file=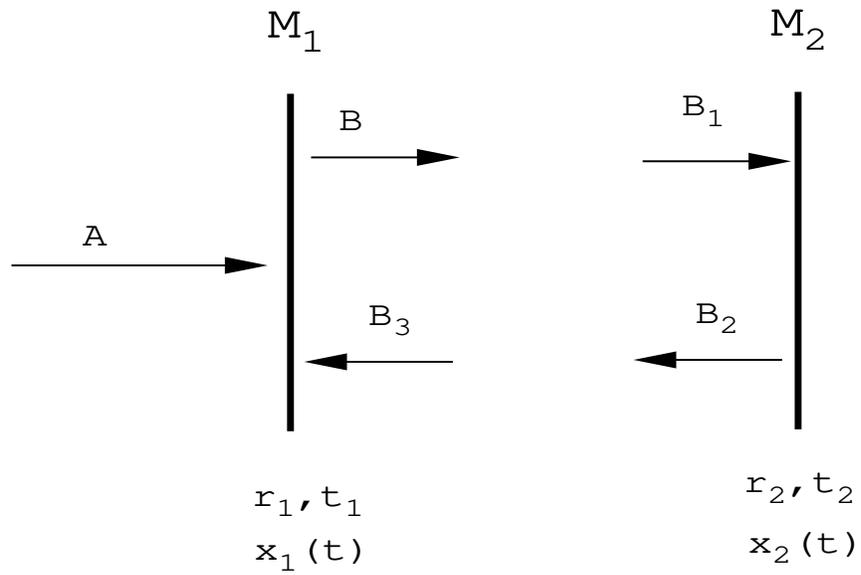,height=3.in,width=4.5in}}
\caption{Schematic diagram of the cavity and the intra-cavity fields.}}
\end{figure}

\begin{figure}{
\centerline{\psfig{file=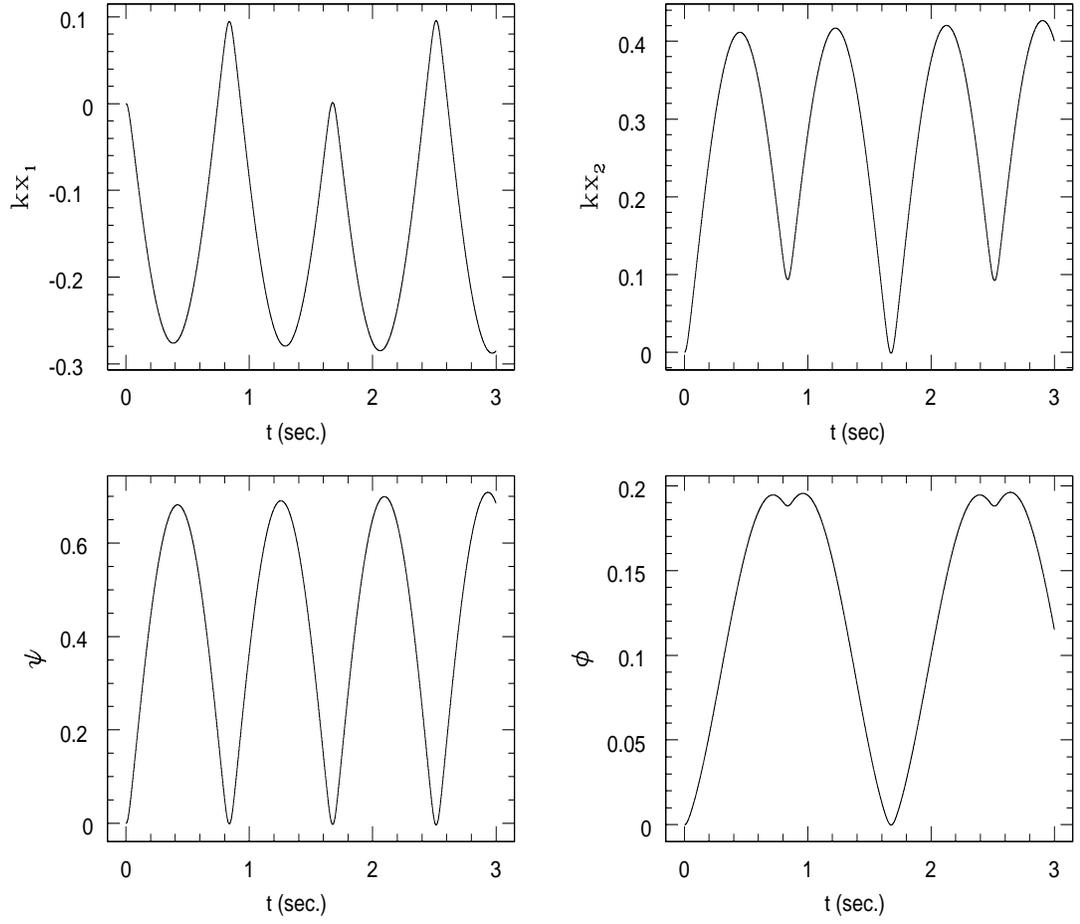,height=7.in,width=4.5in}}
\caption{Motion of the mirrors for the modes $kx_1$, $kx_2$, $\psi$
and $\phi$ as a function of time for input power of 1 kW.}}
\label{figuno2}
\end{figure}

\begin{figure}{
\centerline{\psfig{file=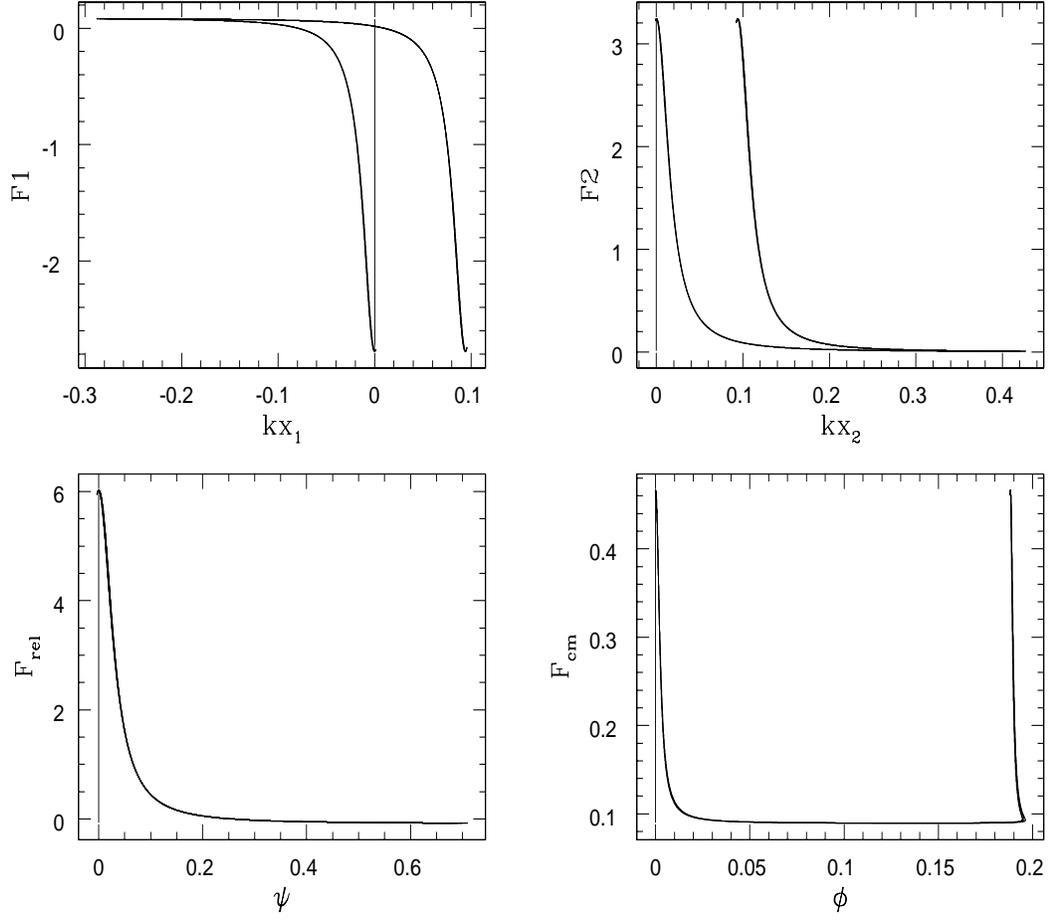,height=7.in,width=4.5in}}
\caption{
Radiation pressure force profiles for the modes $kx_1$, $kx_2$,
$\psi$ and $\phi$ in dimensionless units for input power of 1 kW for one
cycle. The 
conversion factor for converting the dimensionless force into Newtons
is $m\omega^{2}/k$.}}
\label{figuno3}
\end{figure}

\begin{figure}{
\centerline{\psfig{file=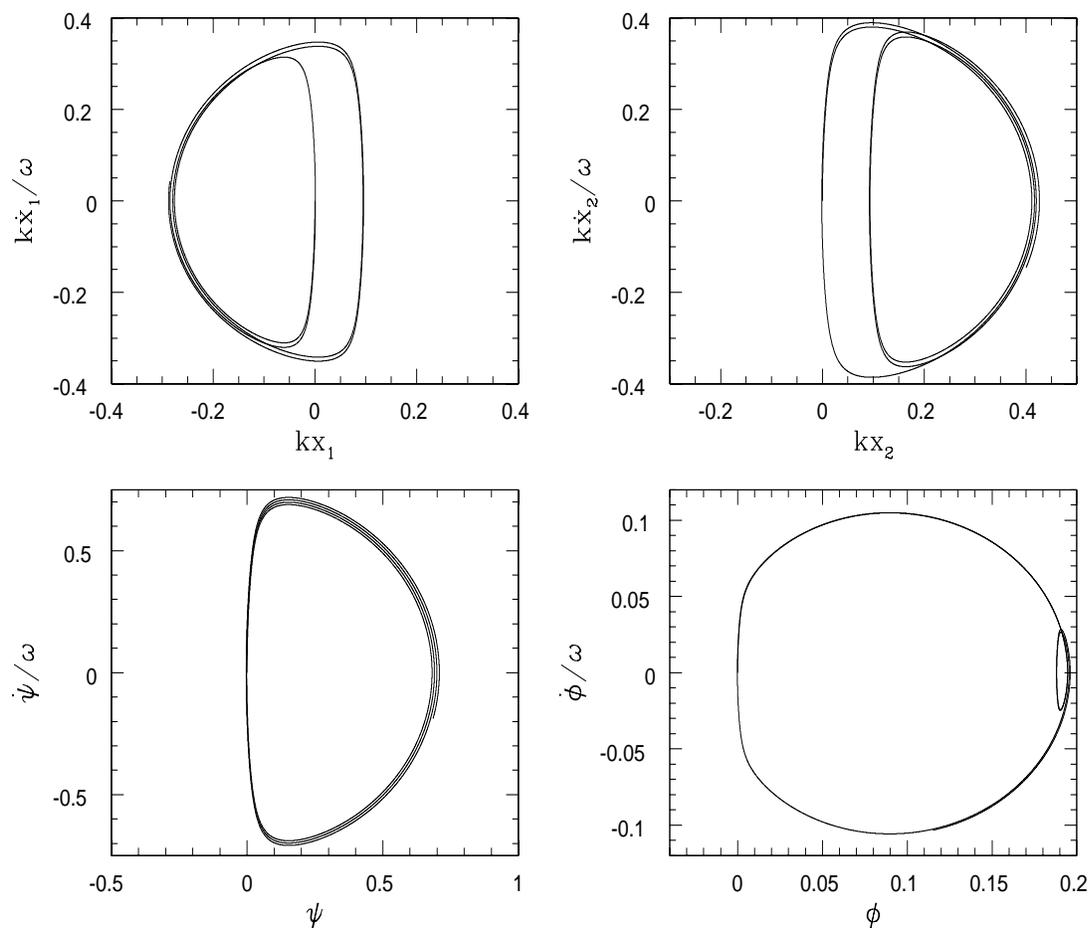,height=7.in,width=4.5in}}
\caption{
Phase space diagrams for $kx_1$, $kx_2$, $\psi$ and $\phi$
for 1 kW of input power and integration time of 3 seconds.}} 
\label{figuno4}
\end{figure}

\begin{figure}{
\centerline{\psfig{file=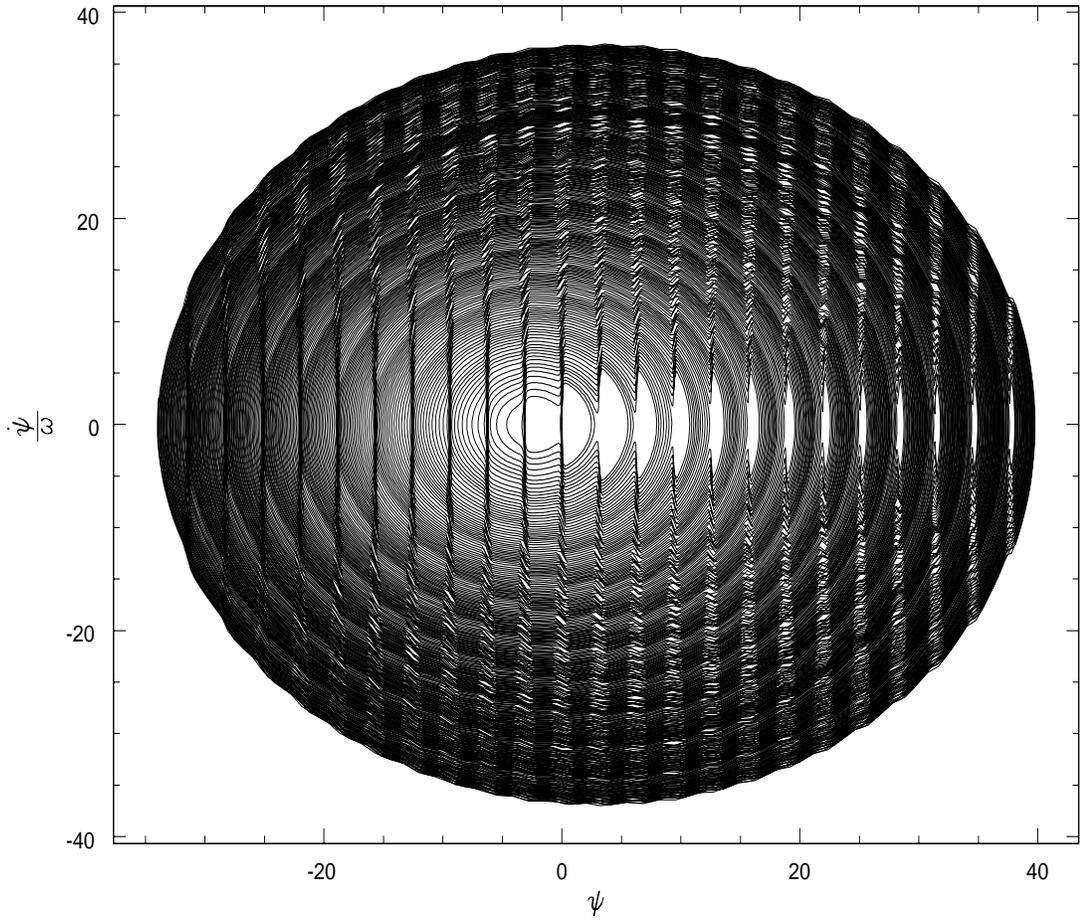,height=7.in,width=4.5in}}
\caption{
Phase space diagram for $\psi$ for the input power of 30
kW and integration time of 500 seconds.}}
\label{figuno5}
\end{figure}

\begin{figure}{
\centerline{\psfig{file=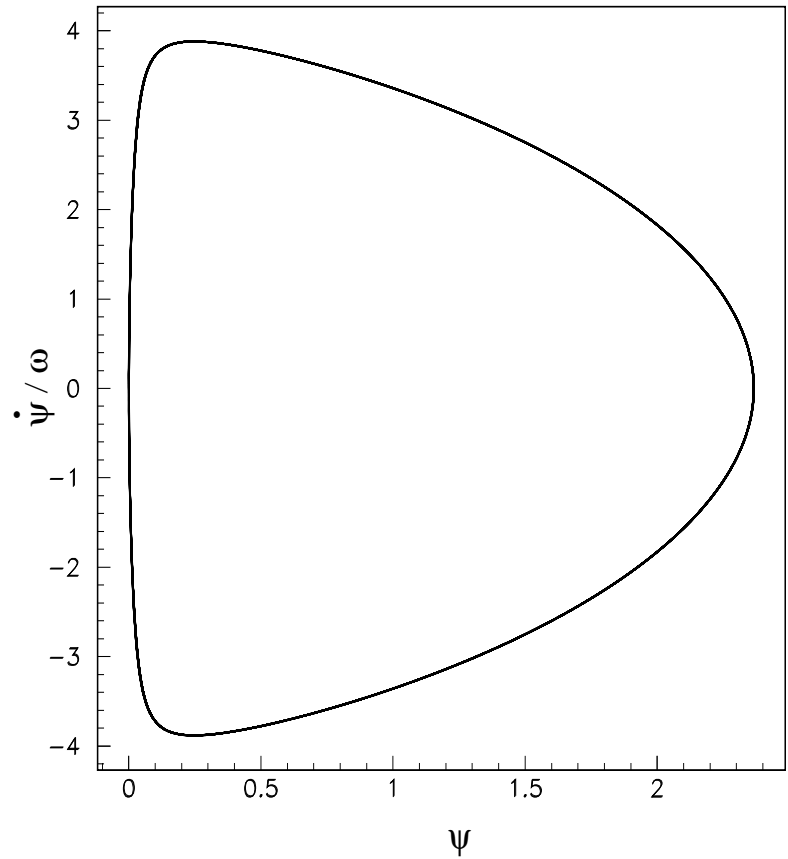,height=5.in,width=4.5in}}
\caption{
Phase space diagram for $\psi$ for the input power of 30
kW,  integration time of 500 seconds and for a short cavity ($L_0=30$ cm). Unlike the long cavity, studied here, there is
no energy gain : the phase-space trajectory remains remarkably stable.}}
\label{figuno5b}
\end{figure}

\begin{figure}{
\centerline{\psfig{file=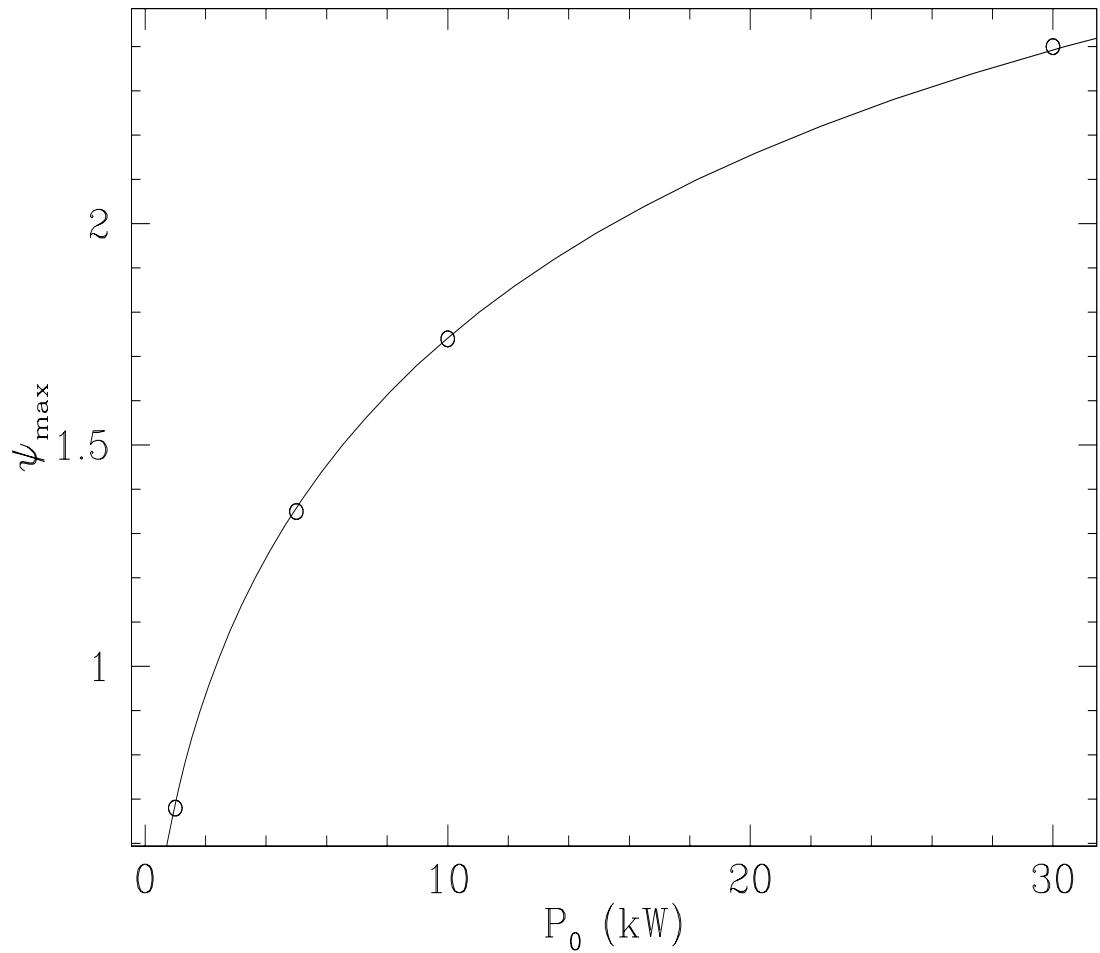,height=7.in,width=4.5in}}
\caption{Comparison of the values of $\psi_{max}$ obtained analytically (smooth
curve) and numerically (open circles) for input powers of 1 kW, 5 kW, 10 kW and 30 kW.}}
\label{figuno6}
\end{figure}

\begin{figure}{
\centerline{\psfig{file=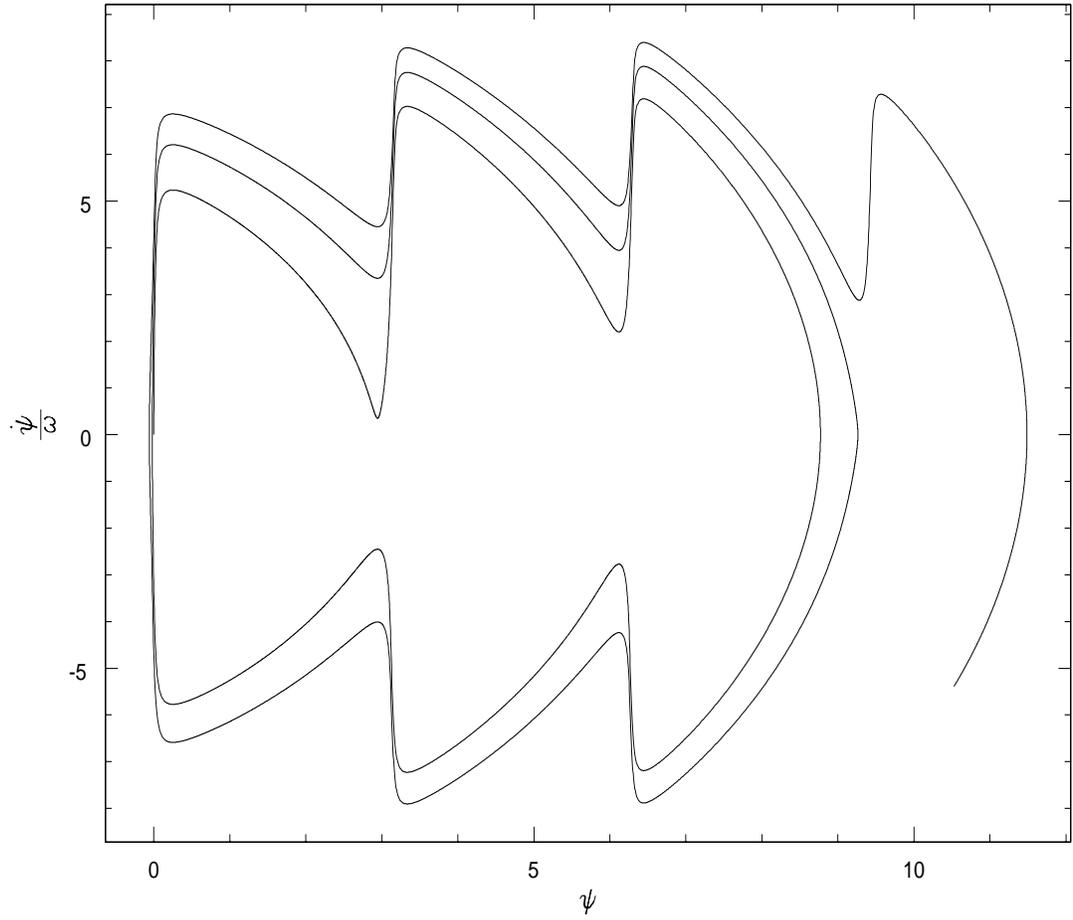,height=7.in,width=4.5in}}
\caption{Phase space diagram for the $\psi$ mode for the input power of 50 kW and integration time of 3 seconds.}}
\label{figuno7}
\end{figure}

\begin{figure}{
\centerline{\psfig{file=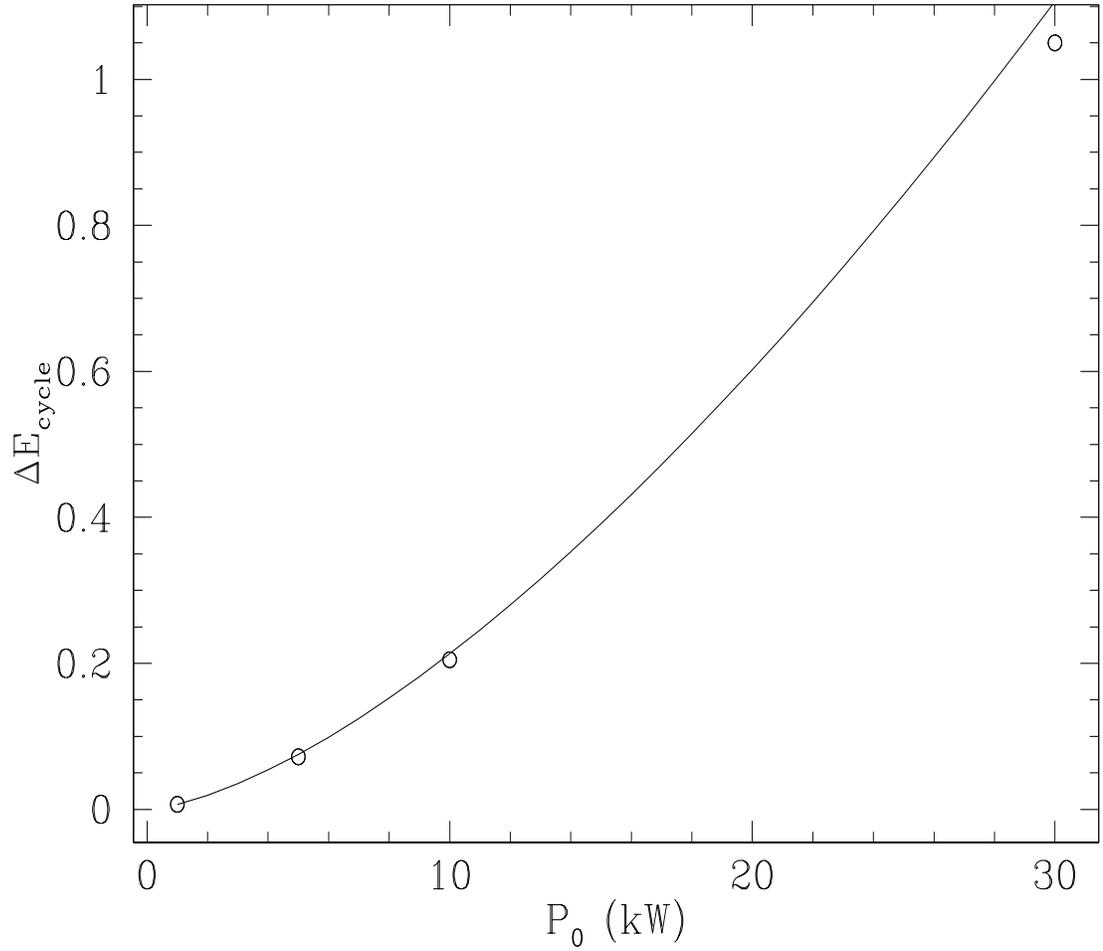,height=7.in,width=4.5in}}
\caption{
The gain in energy per cycle obtained analytically (smooth curve) and
numerically (open circles) for input powers of 1 kW, 5 kW, 10 kW and 30 kW.
To convert the dimensionless energy gain in Joules we multiply
by the factor of $m \omega^2/k^2 \sim 11.2$ pico-Joules.}}
\label{figuno8}
\end{figure}

\begin{figure}{
\centerline{\psfig{file=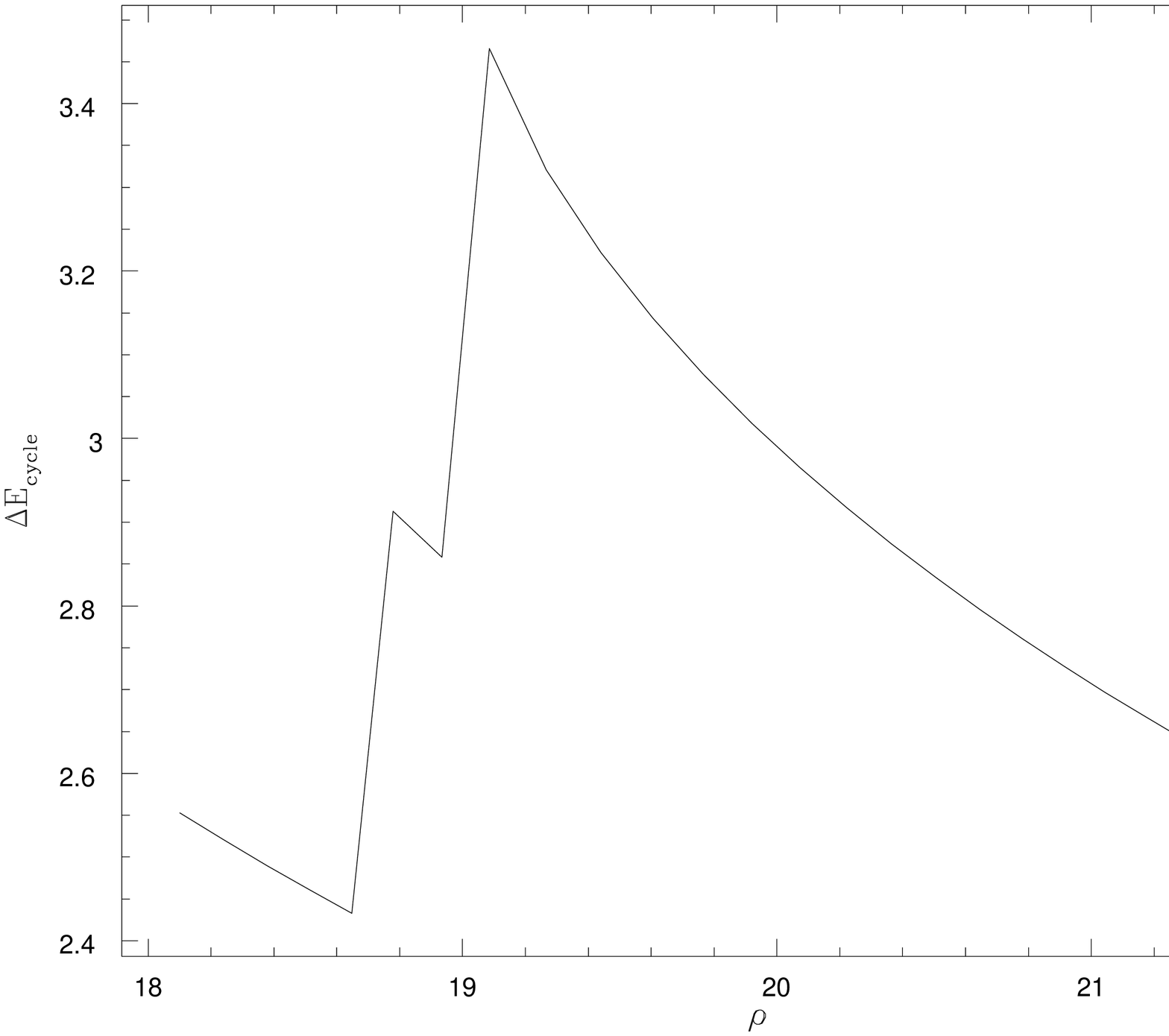,height=7.in,width=4.5in}}
\caption{The dimensionless energy gain per cycle $\Delta E_{cycle}$
 as a function of the 
amplitude of the $\psi$ mode for the input power of 30 kW. The plot is
analytical.}}  
\label{figuno9}
\end{figure}

\begin{figure} {
\centerline{\psfig{file=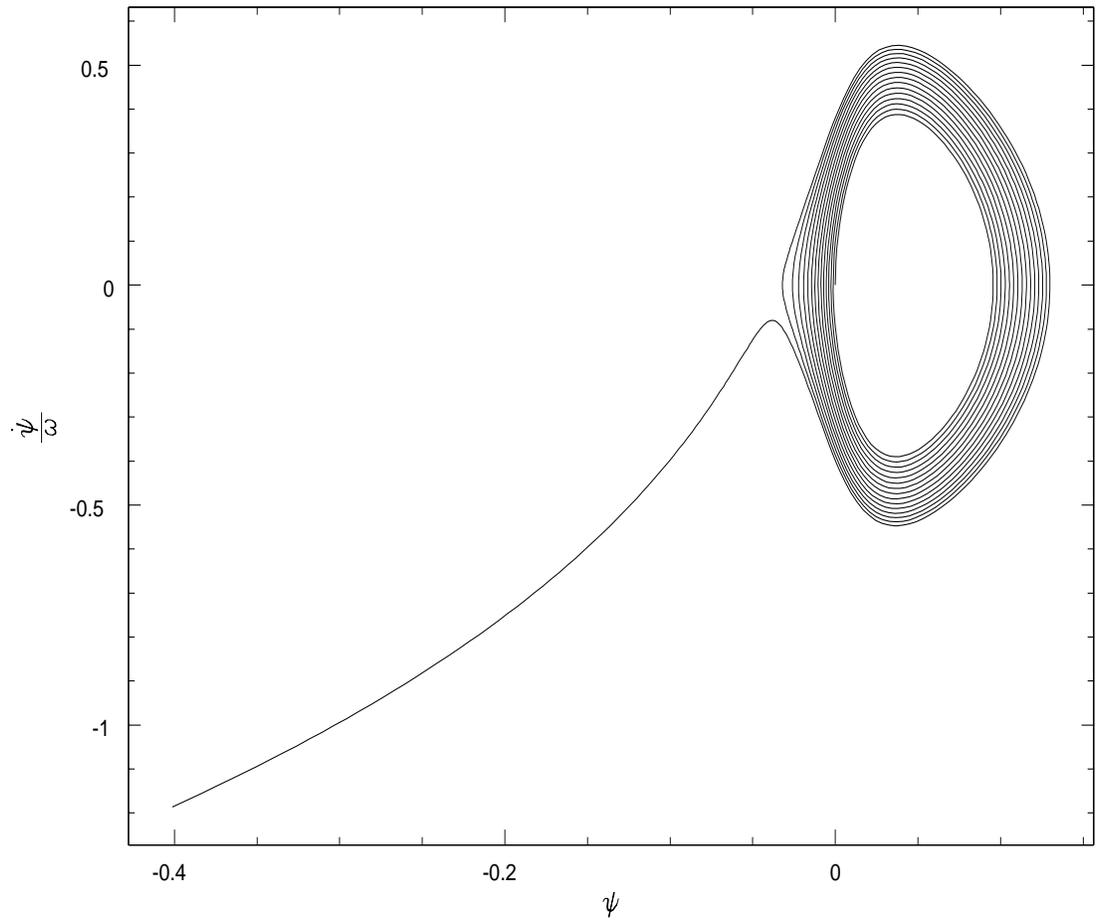,height=7.in,width=4.5in}}
\caption{
The phase space diagram for the equilibrium case for the $\psi$ mode.
The input power is 1 kW and the integration time is 4 seconds.}}
\label{figuno10}
\end{figure}

\end{document}